\documentclass[traditabstract]{aa}
\usepackage{graphicx}
\usepackage{txfonts}
\usepackage{natbib}

\begin{document}

   \title{Non-variable cosmologically distant gamma-ray emitters
    as an imprint of propagation of ultra-high-energy  protons }
    
   \author{A.Yu.~Prosekin \inst{1}
          \and S.R.~Kelner\inst{1}
          \and F.A.~Aharonian \inst{1,2}}

   \institute{Max-Planck-Institut f\"ur Kernphysik,
              Saupfercheckweg 1, D-69117 Heidelberg, Germany\\
              \email{Anton.Prosekin@mpi-hd.mpg.de}
          \and
             Dublin Institute for Advanced Studies, 31 Fitzwilliam Place,
                Dublin 2, Ireland
             }

\date{\today}
\abstract{The acceleration cites of ultra-high-energy (UHE)
 protons  can be traced by the footprint left by these  particles  
 propagating  through cosmic microwave background (CMB) radiation. 
 Secondary electrons produced in  extended  region  of several tens of Mpc 
 emit  their energy via synchrotron radiation  predominantly in the initial 
direction of  the parent protons.  It forms a non-variable and  compact
(almost point-like)  source of high energy gamma rays. The importance
of this effect  is increased for cosmologically distant  objects; because 
of severe energy losses, UHE protons cannot achieve us even 
in the case of  extremely weak intergalactic magnetic fields.  
Moreover,  at high redshifts  the energy conversion from protons to secondary particles
becomes significantly more effective due to the denser and more energetic
CMB in the past. This  increases  the chances of UHE cosmic rays to be 
traced by  the  secondary synchrotron gamma radiation. 
We discuss the energy budget and the redshift dependence of the efficiency 
of energy transfer from UHE protons to synchrotron radiation.
The angular and spectral distributions of  radiation in the
gamma- and X-ray energy bands are calculated and discussed in the context of 
their  detectability by {\it Fermi LAT}  and {\it Chandra} observatories.} 

\keywords{ultrahigh energy cosmic rays -- gamma-ray emission -- X-ray emission
-- point-like source -- high redshift -- propagation}
   \titlerunning{Non-variable cosmologically distant gamma-ray emitters}
   \maketitle

\section{Introduction}
Although there is a little doubt that UHE cosmic rays, $E \geq 10^{19}$eV,
are produced in extragalactic sources, the nature and origin of these objects 
remain highly unknown. In the case of  very weak  intergalactic magnetic fields (IGMF), 
the initial directions of these particles is  only  moderately  entangled, and thus 
the information about  their acceleration sites  can be partly  preserved.
The  quantitative  treatment of  propagation of UHE cosmic rays is limited because of 
large (orders of magnitude) uncertainties of relevant  model parameters. 
Especially it concerns the IGMF.  The current measurements  
give only upper limit $B\sim10^{-9}$ G for  the component parallel to the line of sight \citep{Ryu1998,Blasi1999};  the lower bound is estimated as small  
as  of $B \sim10^{-16}$~G (see \cite{Taylor2011} and references therein) 
from the non-observations of GeV gamma-ray counterparts from distant 
TeV blazars.  The large scale structure of IGMF  is  also very uncertain;  
the attempts to model usually lead to quite different results \citep{Dolag2005,Sigl2004}. These uncertainties significantly limit the potential of the so-called cosmic ray astronomy.
Moreover, even in the case of  very weak magnetic fields, the  cosmic ray "horizon"  of UHE cosmic ray sources through direct detection   of these particles is limited  by interactions with CMBR.  For highest energy  particles, $E \sim 10^{20}$eV,  it   does not exceed 100 Mpc.  At lower energies cosmic rays can propagate to significantly  larger distances, but they are dramatically deflected by galactic magnetic fields. These two factors significantly limit the potential of the  so-called {\em proton astronomy}.    

An important information about the acceleration  sites of  UHE protons 
is contained in gamma rays produced  during  propagation of protons 
through the CMBR.  In the case of very low intergalactic magnetic field,
the secondary products  of $p\gamma$ interactions  initiate electromagnetic 
cascades in CMBR. 
The  effective  development  of  such cascades  is determined by the  
dominance of Compton energy losses over the synchrotron losses of  secondary 
electrons. Generally,  this condition is quite  relaxed;  
it requires IGMF   weaker than  $10^{-10}$G. However, the second condition 
of detection of cascade gamma rays in the direction of   parent UHE protons requires 
much weaker magnetic fields, smaller than $10^{-15}$G. Otherwise the relatively
low energy electrons, and consequently the GeV and TeV gamma-ray emission,  
will be immediately isotropised.   Formally, such week intergalactic 
magnetic fields cannot be excluded. Moreover, recently it has been argued that very high energy tails of  gamma-ray spectra  of distant TeV  blazars might be contributed 
by cascades  initiated by  UHE protons in  intergalactic medium with magnetic 
field as small as $10^{-15}$G \citep{Essey2011}.  

Remarkably,  even in the case of strong IGMF, we might expect physically 
extended  (but  looking  point-like)  gamma-ray sources formed 
around  production sites of UHE cosmic rays \citep{Gabici2005,APK2010,Kotera2011}.
In this case 
gamma rays are produced  through synchrotron radiation of secondary electrons 
from photomeson  interactions. Protons lose significant fraction of their energy 
on  scales of several tens of Mpc.  If the magnetic field in these regions located 
at relatively small redshifts, $z \ll 1$, are 
significantly large, $B \geq 10^{-9}$G  
(so the energy of secondary electrons is released in 
the gamma-ray band), but smaller than $10^{-7}$G (thus the initial directions of electrons are not dramatically changed),  the resulting  gamma-ray synchrotron sources
can be quite compact with an angular size as small as 0.1$^\circ$. 

For cosmologically distant sources embedded in denser 
($n_{\rm ph} \propto (1+z)^3$) and more 
energetic  ($T \propto 1+z$) CMBR,  protons 
lose  their energy on  distances  significantly smaller than $100$ Mpc. In the presence
of magnetic field  of  strength $B=10^{-9}-10^{-8}$~G the high energy electrons
produced in photomeson processes intensively emit synchrotron radiation in
GeV range of energies. The electrons radiate most of their energy on
almost rectilinear part of the path that gives, along with the small deflection
of protons, collimated beam of gamma rays pointing to the acceleration site
of UHE protons.  Note that the considerable contribution to the high energy electrons
is provided by the gamma rays produced in decay of mesons. The interaction
of these gamma rays with cosmic radio background radiation occurs in the
regime when the most of the energy goes to one of the component of the created
electron-positron pair. 

The   probability  of detection of  cosmic ray sources 
via synchrotron radiation of secondary  electrons 
strongly depends on the maximum energy of accelerated 
protons.  In the case of  sources with redshift $z \ll 1$,
the interaction of protons with CMBR is effective only when the 
proton spectrum extends to  $10^{20}$eV and  beyond. Since   
acceleration of protons to such high energies can be  
realized only at a unique combination of a few principal parameters,
in particular  the linear size of the source, the strength of the magnetic field, and 
the Lorentz factor of the the bulk motion (\cite{FAA2002}),    the number of sources 
of $10^{20}$eV cosmic rays can be quite limited compared to sources accelerating particles to  $E_{\rm max} \sim 10^{19}$eV.   Since  protons at high redshifts interact with 
denser and more energetic photons of CMBR,  the  requirement to $E_{\rm max}$ 
is  significantly relaxed, thus one should expect dramatic increase of the number of 
UHE cosmic ray sources surrounded by gamma-ray halos. 
Furthermore, at high redshifts the interaction of protons with
CMBR  via pair production (Berhe-Heitler) process begins to play an important role
and a  considerable part of proton energy is converted to less energetic electrons
compared to  electrons  generated in photomesons  reactions. 
Appearance of dense
halos of  Bethe-Heitler   electrons around  the source  at presence of 
magnetic field of the order of  $B\sim10^{-6}$~G 
(comparable to the field  of clusters of galaxies) 
results in  radiation dominated by synchrotron X-rays. 

In this paper we study the energy and angular distribution of
the synchrotron gamma-ray emission from cosmologically distant sources. The
calculations are based on the approach developed in  our previous work (\cite{APK2010}) combined with cosmological effects on propagation of gamma
rays and protons.  The space in the vicinity of the source, where all
relevant processes occur, can be considered as conventional one. Therefore,
taking into account  denser ambient
radiation at high redshifts, we can apply the developed formalism to calculation
of the distribution function
of gamma rays in the vicinity of the source. Using this function, we can
easily obtain the distribution function of the observed radiation which has
propagated cosmological distance (see Appendix \ref{app}).

\section{Energy budget}
\begin{figure}
\begin{center}
\includegraphics[width=0.5\textwidth,angle=0]{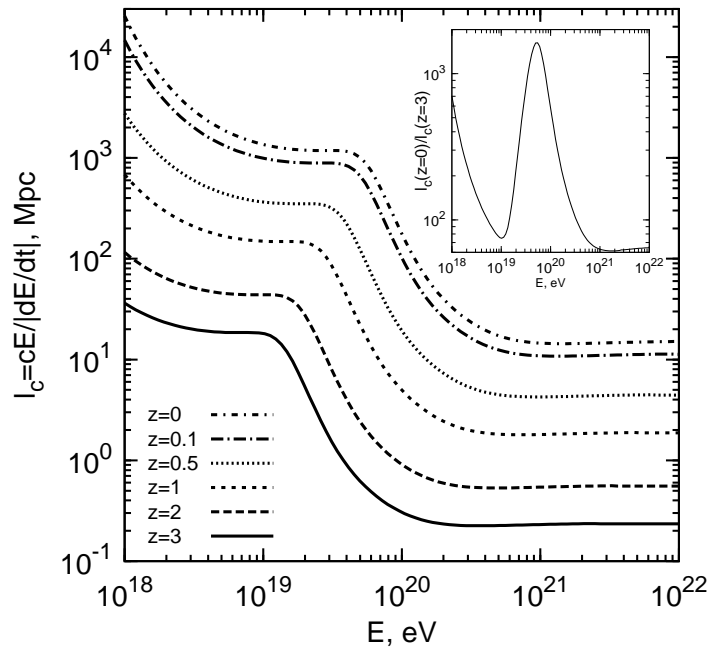}
\caption{\label{fig1}The cooling length of protons in the intergalactic medium
due to interactions with
photons of CMBR at different redshift. The ratio of cooling lengths  at $z=0$
and $z=3$ are  shown in the inset.}
\end{center}
\end{figure}

An accelerator of UHE protons located at high redshift has appreciably
different conditions of ambient media as compared to the ones in the
nearby Universe. The photon field of CMB is denser and more energetic due
to cosmological expansion. The increase of density  by factor of $(1+z)^3$
leads to more frequent
interactions of UHE protons with CMB that intensifies the energy
loss rate. The average energy of photons is also increased  by factor of $(1+z)$ that decreases the threshold energy of the interactions
for protons.
The  energy loss rate of protons due to interaction with CMB 
at the epoch of redshift $z$ is expressed in terms of the present loss rate
$\beta(E)=-\frac{1}{E}\frac{dE}{dt}$ as  
\begin{equation}
\beta(E,z)=(1+z)^{3}\beta_0((1+z)E).
\end{equation}
It is convenient to present this relation in terms of the cooling length
which is shown in Fig.~\ref{fig1}. The cubic dependence on redshift leads
to considerable decrease of the cooling length. In particular, for the constant
losses at highest energies, $E \geq 10^{21}$eV, 
 it is reduced from  $\approx 15$ Mpc at the present epoch
to $\approx 0.2$ Mpc at $z=3$. The plateau of constant losses itself extends
to lower energies due to the energy shift. At lower energies, the combination of the 
effects related to the  energy shift  and the increase of density jointly 
results in reduction of cooling length by a factor larger than $(1+z)^3$. 
Indeed, as it is seen from the inset in Fig.~\ref{fig1}
the reduction of the proton cooling length can be an 
order of magnitude larger. The peaks show 
domains where energy loss at $z=3$ is the most intensive relative to the
case of present epoch. If the energy cutoff in the initial 
proton spectrum falls into this domain, the advantage of energy extraction
at cosmological distances becomes more significant. Moreover,
 at  $E=10^{18}$eV,  the  cooling length  is reduced  from $\geq 10^4$ Mpc at $z=0$ 
to tens of Mpc  at  $z=3$.  It is interesting to note  that the energy
loss rate of protons of energy  $E<10^{19}$ eV at  $z=3$  is comparable to the loss rate 
of $E \geq 10^{20}$eV  protons  at the present epoch. 
\begin{figure*}
\begin{center}
\mbox{\includegraphics[width=0.5\textwidth,angle=0]{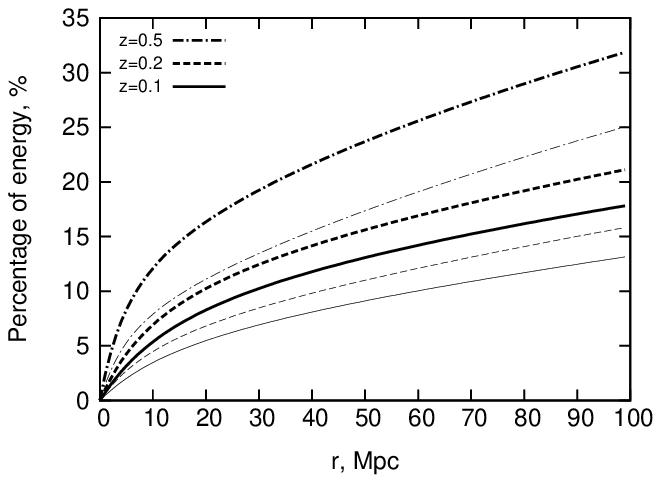}
\includegraphics[width=0.5\textwidth,angle=0]{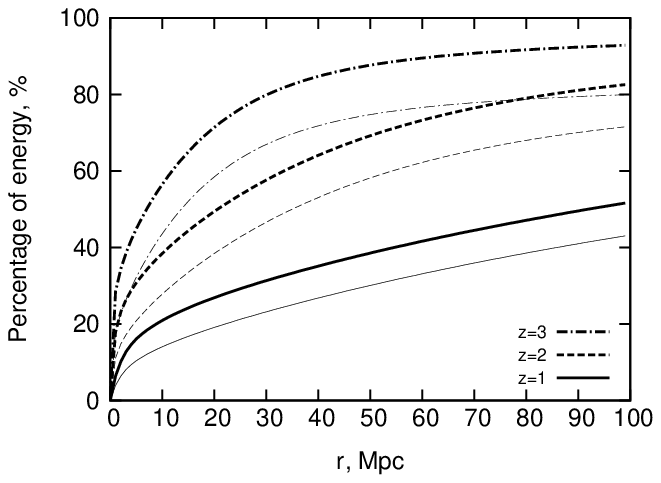}}
\caption{\label{fig2} The fraction  of the initial energy of protons with
$E>10^{18}$ eV  lost  (thick lines) and converted to
the energy of electrons (thin lines) at the distance $r$ from the source.
The  injected proton spectrum is  assumed power-law with an exponential cutoff  
$J_{p}(E)=J_{0}E^{-2}\exp(-E/E_0)$, with $E_0=3\cdot10^{20}$ eV.}
\end{center}
\end{figure*}

Fig.~\ref{fig2} describes the evolution of proton energy losses and the efficiency of their
conversion to the electron component of secondaries with the distance to  the
source at different redshifts.  The electron component includes the electrons
produced through all channels under consideration: pair production by protons,
decay of charged mesons produced in photomeson processes and pair production
by gamma rays produced from decay of neutral mesons. As mentioned above,
in the latter process almost all energy of the photon goes to the energy of one
of the electrons. Therefore the gamma rays can be treated as electrons.
Then the energy taken away by neutrinos is the difference between the
energy lost by protons and the energy converted to electrons.
As can be seen from Fig.~\ref{fig2},  the protons with initial 
energies $E>10^{18}$ lose only $18\%$ of
their total energy after passing $r=100$  Mpc at the redshift $z=0.1$, whereas at redshift
$z=3$ the same protons lose  $93\%$ of the available energy  already at $r=70$ Mpc,  where the saturation begins. It can be explained by the fact that the cooling length of protons in the range
$E>10^{18}$ eV do not exceed tens of Mpc at $z=3$, while for redshifts $z \ll  1$, 
the  protons have the cooling length of cosmological scales
relative to pair production process. As the contribution of pair production
process increases, the share of energy lost by protons that goes to electrons
increases  from $74\%$ at $z=0.1$ to $86\%$ at $z=3$ 

\begin{figure*}
\begin{center}
\mbox{\includegraphics[width=0.5\textwidth,angle=0]{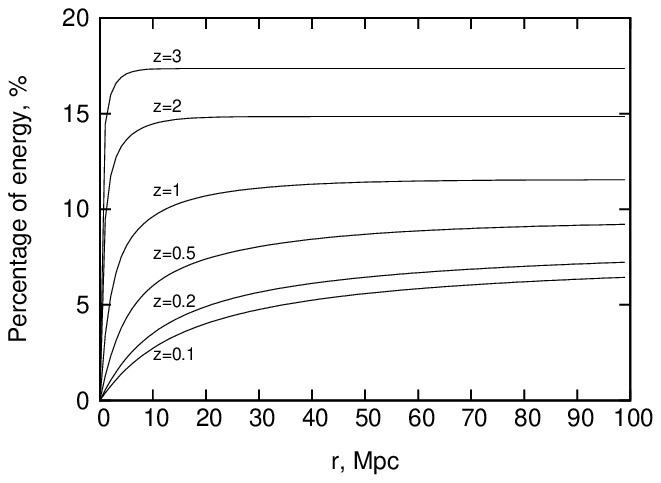}
\includegraphics[width=0.5\textwidth,angle=0]{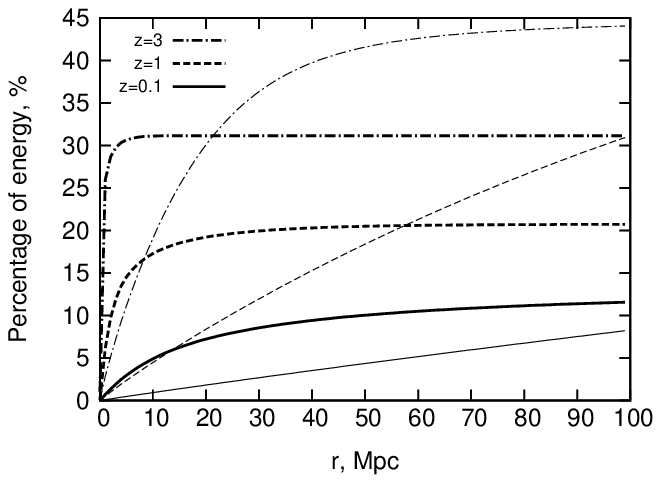}}
\caption{\label{fig3}Left panel: the fraction  of the initial energy of
protons with
energy $E>10^{18}$ eV converted to the energy of electrons from \textit{photomeson
production}  at  distance $r$ from the source. Right panel: the percentage
of the initial energy of protons with energy $E>10^{19}$ eV converted to
the energy of electrons from photomeson production (thick lines) and electrons
from pair production (thin lines) at the distance $r$ from the
source.}
\end{center}
\end{figure*}

The  fraction  of the initial energy of the protons with $E>10^{18}$ eV that
can be converted to the electrons generated at photomeson processes depends
strongly  on the position of the cutoff energy. Left panel of Fig.~\ref{fig3}
shows that the
efficiency of extraction of proton energy and its transfer to this component
of electrons grows with the
redshift, and at $z=3$ all available for conversion energy is transferred at first $5$ Mpc.
The fraction  of proton energy taken away by neutrinos in photomeson processes
is $42\%$ independently of redshift. Due to more
energetic photons of CMBR, the  threshold of photomeson interactions is shifted to
lower energies that leads to the increase of extracted energy from $6\%$
to $17\%$ (see Fig.~\ref{fig3}). However, as pair production begins to play
a significant role at high redshifts the share of the electrons generated
in  photomeson processes  is decreased,  from $49\,\%$
to $22\,\%$ at the distance $r=100$ Mpc. 

The importance of the shift of the reaction threshold on 
production of electrons  becomes more obvious
if the protons with energies close to threshold and higher are taken into
account. The percentage
of initial energy of protons with energies $E>10^{19}$ eV which is converted
to the energy of electrons is presented in  the right panel
of Fig.~\ref{fig3}.   While at $z=0.1$ the production of electrons due to
photomeson processes  dominate at all distances smaller than $100$ Mpc, at
high redshifts the pair production  prevails.

\begin{figure*}
\begin{center}
\mbox{\includegraphics[width=0.5\textwidth,angle=0]{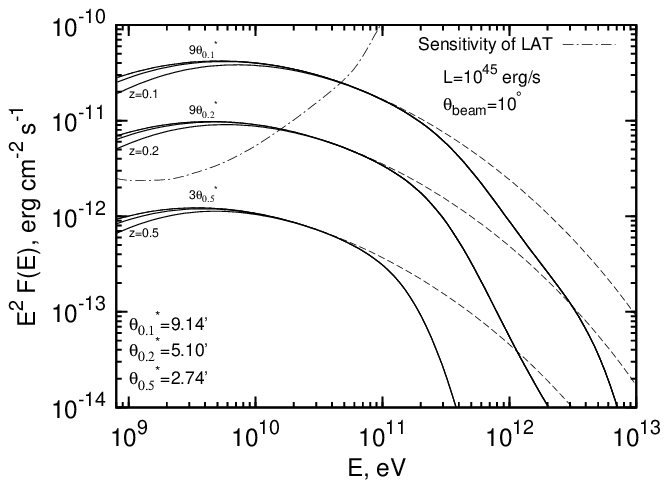}
\includegraphics[width=0.5\textwidth,angle=0]{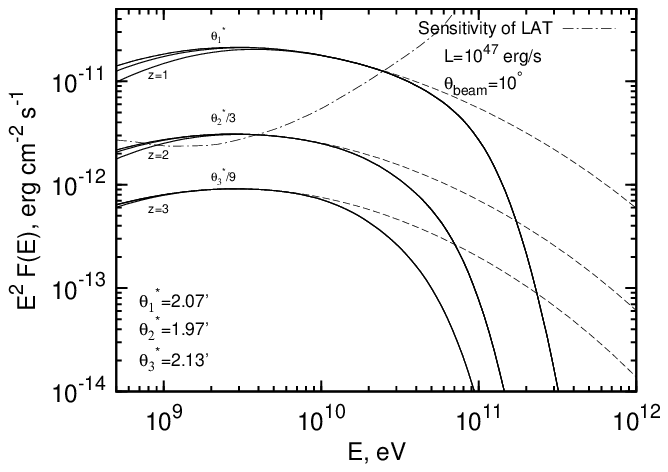}}
\caption{\label{fig4} Spectral energy distribution of gamma rays observed
within different angles in the direction to the source with EBL absorbtion
(thick lines) and without one (dashed lines) indicated for maximum angles.
$\theta^{*}_{z}$ is  apparent angular size of the region with
radius $0.5$ Mpc from the distance with redshift $z$. The angles are specified
in the units of corresponding $\theta_{z}^{*}$ and differ from each other
by factor of 3. The maximum angles for every redshift, the total power
of injection of protons $L$  and the beaming angle $\theta_{beam}$ are indicated.}
\end{center}
\end{figure*}

\section{Gamma-ray source}

Having the highest energy among secondaries, the electrons generated in photomeson
processes emit almost all their energy via synchrotron radiation. Therefore,  
because of the shift of the threshold of photomeson interactions, 
the intensity of synchrotron gamma rays is increased with the redshift of the source.
On the other hand,  the reduction of  cooling length of protons 
results in  reduction of the apparent angular size of the region emitting synchrotron radiation,  in addition to the diminution because of geometrical factor.

Fig.~\ref{fig4} shows spectral energy distribution  of synchrotron gamma
rays observed within different polar angle in the direction to the source
of protons. The geometry of expanding Universe leads to more sophisticated
dependence of apparent angular size on distance 
compared to the  $\sim 1/r$  dependence for the local Universe. It is useful
to define a reference angle $\theta_{z}^{*}$ which is equal to the angular
size of the region with the radius $r=0.5$ Mpc located at the redshift $z$. 
Expressing the angles in units of 
corresponding reference angles, we can compare the  
regions located at different redshifts eliminating the geometrical
factor.  As expected, at large redshifts  the reference angle increases 
with $z$ (see Appendix \ref{app}, Fig.~\ref{fig13}).  
In Fig.~\ref{fig4} for each redshift  the fluxes are presented within three polar angles which
differ from each other by factor
of 3 and are expressed in the units of corresponding reference angle. The maximal
angle indicated in plots  is the polar angle within which the total flux
is observed. Comparison of the maximal angles in  units
of reference angles shows  a  tendency of  decreasing the  angular size of
the region of secondary synchrotron radiation with redshift, 
from $9\theta^{*}_{0.1}$ to $\theta^{*}_{3}/9$. 

The interaction of synchrotron gamma rays with extragalactic background light
(EBL) leads to considerable attenuation of the flux. 
 As it is seen from Fig.~\ref{fig4} the intergalactic absorption becomes 
 substantial, depending on the distance to the source, 
 from TeV energies down to tens of GeV energy range.  For calculation of the
absorbtion of gamma rays, the model of EBL developed in \cite{Franceschini2008} have been applied. 

At high redshifts, the 
electrons  from  photomeson  reactions  are produced close to the acceleration
cites of protons. In this case a considerable part of emitting electrons
might be found in much stronger magnetic field compared to the average 
intergalactic  field. The energy spectrum of synchortron radiation of secondary 
electrons is shifted linearly with change of the strength
of magnetic field to keep the ratio $E_{\gamma}/B$ constant (see  \cite{APK2010}).
Therefore the  increase of the strength of magnetic field   leads to the shift  of  radiation spectrum towards  higher energies; if the radiation extends to TeV energies, the intergalactic absorption  becomes quite  violent;  this results in the dissipation
of almost the entire radiation. The absorbed gamma rays initiate cascades which contribute to the diffuse extragalactic background radiation.    
 
Keeping in mind the sensitivity of instruments such as {\it Fermi}, the detection
of the collimated synchrotron radiation from sources located at high redshifts
is possible only in case of very powerful AGN. The anisotropic injection
of UHE cosmic rays allows to reduce the required power of source. Until deflection
of protons is smaller than the angle of a jet, there is no difference between the 
spherically symmetric and anisotropic case. In both cases the  observer will
see identical pictures. For the  given power of injection the existence
of anisotropy
with the opening angle $\theta_{beam}$ of jet
means the increase of the flux of gamma rays by the factor  $4/\theta^2_{beam}$
compared to the  spherically symmetric case. The fluxes of synchrotron
gamma rays
in Fig.~\ref{fig4} are calculated for the case of $\theta_{beam}=10^{\circ}$.
The deflection of protons is smaller than this value of the opening angle
while protons produce secondary particles. 
It is  seen from  Fig.~\ref{fig4} that at the power of proton injection 
$L=10^{45}-10^{47}$erg/s, the fluxes of gamma rays can exceed the sensitivity 
(minimum detectable fluxes) of {\it Fermi LAT} \citep{Atwood2009}
The resolution of the {\it Fermi LAT}
varies in dependence of photon energy from $42'$ at $E=10^{9}$ eV  to $4.2'$
at $E=10^{11}$ eV \citep{Atwood2009}. Already for the source located at $z=0.2$ 
the angular size is $46'$, thus the source of gamma rays would be seen
as extended one only if $z<0.2$.

\section{X-ray  emission}
\begin{figure}
\begin{center}
\includegraphics[width=0.5\textwidth,angle=0]{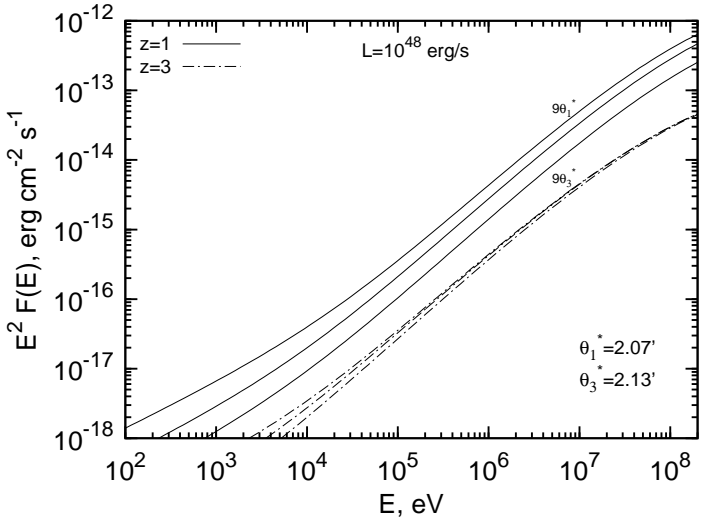}
\caption{\label{fig5}Energy flux distribution of gamma- and X-rays
observed within different angles in the direction of the source for
the case of homogeneous magnetic field $B=1$ nG. Other parameters  
are  the same as in previous figures.}
\end{center}
\end{figure}

\begin{figure*}
\begin{center}
\mbox{\includegraphics[width=0.5\textwidth,angle=0]{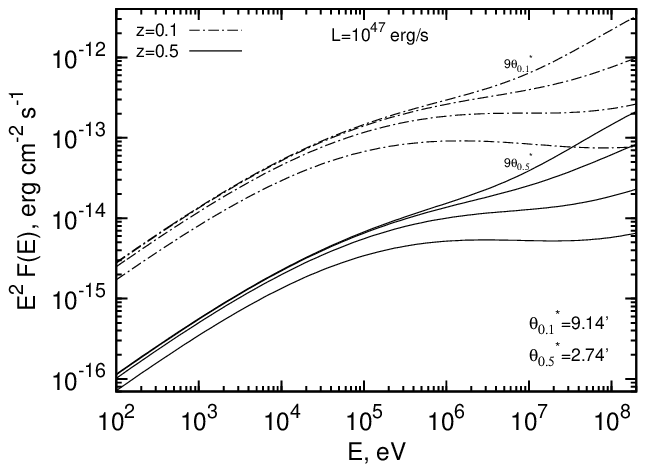}
\includegraphics[width=0.5\textwidth,angle=0]{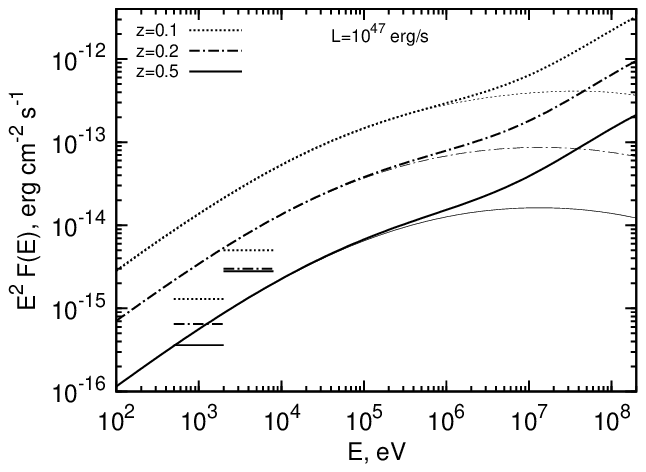}}
\caption{\label{fig6} Energy flux distribution of gamma- and X-rays observed
within different angles in the direction of the source (left panel) and within
maximum angle (right panel) for the case of two-band magnetic field. For
the right  panel, the total radiation from photomeson and pair production
electrons (upper lines) and the radiation from
pair production electrons (lower lines) are indicated. Horizontal segments
present the {\it Chandra} sensitivity for the corresponding maximum angles of
observation.  The other parameters  are the same as in  previous figures.}
\end{center}
\end{figure*}

\begin{figure*}
\begin{center}
\mbox{\includegraphics[width=0.5\textwidth,angle=0]{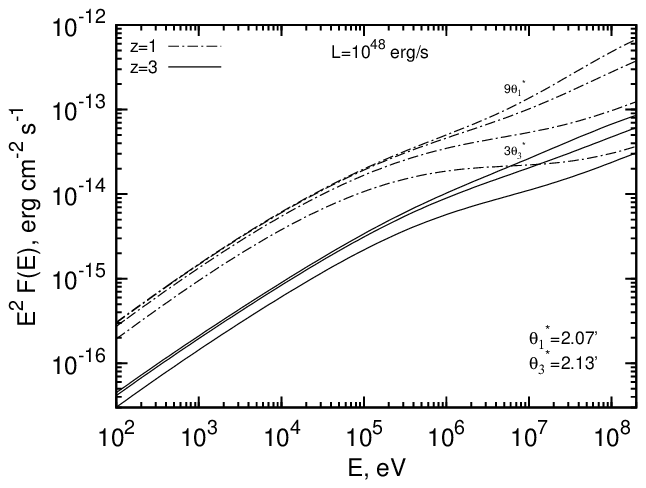}
\includegraphics[width=0.5\textwidth,angle=0]{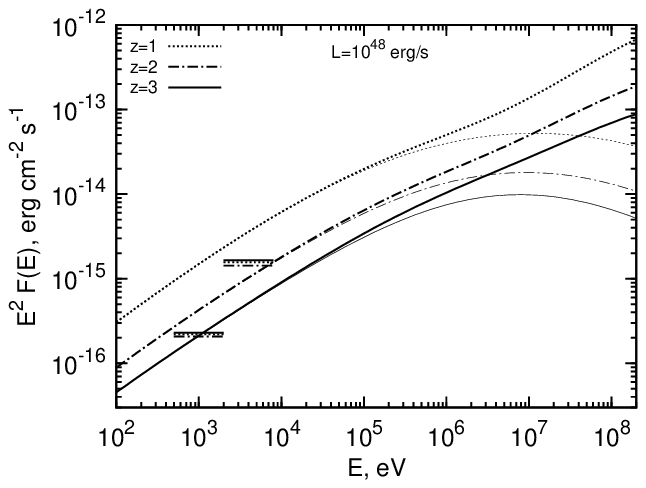}}
\caption{\label{fig7} The same as  Fig.~\ref{fig6} but   for  $z=1,2,3$ and
the total power of injection of protons $L=10^{48}$ erg/s .}
\end{center}
\end{figure*}
Protons lose part of their energy via pair production. The pair-produced electrons
have lower energies compared to  the  electrons  produced in photomeson
processes and emit synchrotron radiation in the lower band
of spectrum. At high redshifts the mean free path of protons relative to
pair production process is decreased and generated electrons are located
in more compact region. Moreover, the fraction  of energy of protons converted
to the secondary electron component is increased and reaches to 
$68\%$ at $z=3$ (see Fig.~\ref{fig2}, \ref{fig3}).

Synchrotron radiation for the strength of IGMF of $B=~10^{-9}$~G and inverse
Compton (IC) scattering give equal contribution to electron losses at energy
$E\approx3\cdot10^{18}$ eV. At lower energies electrons lose their energy
predominantly through IC scattering. Scattering is carried out at Klein-Nishina
regime and almost all energy is transferred to photon. In turn the high energy
photon produce electron-positron pair due to interaction with CMBR and most
of the energy goes to one component of the pair which again suffer IC scattering.
This process can be considered as alternation of the particle state
with gradual
reduction of energy. The electromagnetic cascade produces a huge halo of
gamma rays with energy in GeV region. Region of synchrotron radiation is
more compact, and electrons radiate it  at the place of their generation.
In spite of this the region is still quite extended as can be seen from angular
distribution on Fig.~\ref{fig5}. The flux of synchrotron radiation drops
at low energies and becomes very small in X-ray region.  The
calculation of fluxes presented on Fig.~\ref{fig5} takes into account only
homogeneous IGMF with strength $B=10^{-9}$G. However, the strength of magnetic
fields close to the accelerator can be much higher. To take this into account,
we consider the following spatial distribution for IGMF magnetic field:
\begin{equation}
B=B_{cl}\left(\frac{r_0}{r+r_0} \right)^3+B_0,
\end{equation}
where 
$B_{cl}=10^{-6}{\rm G}$ is magnetic field in the cluster of galaxies with
characteristic size of $ r_0=1$~Mpc, $B_0=10^{-9}$~G is IGMF.
Such a  dependence of the magnetic field has  been
chosen to have a dipole behavior at large distances. For this case the energy
and angular distributions of gamma- and X-rays are
presented on left panels  of Fig.~\ref{fig6} and \ref{fig7}. This magnetic field   
gives considerable increase of the fluxes of synchrotron radiation
at low energies which is generated at small region close to the source and,
therefore,
has narrow angular distribution. Right panels shows that the contribution
of electrons produced in pair production process (lower lines) is dominant
in X-ray region, whereas gamma rays are generated predominantly by electrons
produced in photomeson processes. Horizontal segments indicate sensitivity
of {\it Chandra} \citep{Lehmer2005,Romano2008} for corresponding maximum angles of observation.
As can be seen the low energy part of X-rays ($0.5-2$ keV) is detectable
for specified power of injection of protons, whereas the radiation
at  higher energies can be missing for sources located at high redshifts
since {\it Chandra} is less sensitive in the range of energies $2-8$ keV.

As follows from Fig.~\ref{fig6} and \ref{fig7},  the electrons produced in photomeson
processes  provide the main contribution to synchrotron gamma rays, although
the most of  energy lost by protons is contained in the low energy component
of secondary electrons from the pair production process. 
The radiation of this component can be detected only in the case of a quite large, 
$B_{cl} \simeq 10^{-6}$ G   magnetic field around the 
source. This can happen if the source is located inside a cluster of galaxies
with a typical size of 1 Mpc.  Nevertheless, even in the case
of intensive pair production that takes place in cosmologically distant objects,  
only a small  fraction  of pairs  is  produced  in  the relatively 
compact region with  a linear size of  $\sim 1$ Mpc (see Fig.~\ref{fig2});  the 
most of energy of protons is converted to the energy of extended
gamma ray halo.
\section{Summary}
High energy gamma rays are  unavoidably formed around the sources of UHE
cosmic rays because of synchrotron radiation of secondary electrons produced
at interactions of highest energy protons with the cosmic microwave background
radiation. In our previous paper (\cite{APK2010}) we have shown that even for
relatively large intergalactic magnetic fields in the neighborhood of UHE
cosmic ray accelerators, $B \sim 10^{-7}-10^{-9}$~G, these process lead
to  formation
of high-energy point-like gamma-ray sources. Because of relatively weak
gamma ray signals, the chances of detection of such sources are higher for
objects located in the nearby Universe, namely at distances less than 100 Mpc.
Since the efficiency of conversion of the proton energy to secondary
gamma rays is dramatically reduced at protons energies $E \le 10^{20}$ eV, 
one may hope to detect gamma rays only from extreme objects accelerating protons
to energies $10^{20}$~eV and beyond. Given the fact that this requirement can be
satisfied only in the case of  unique combination of parameters, as well as the
limited volume of the $\leq 100$~Mpc region, 
realistically one can expect not a very large number of such sources. One can gain 
a lot if extends the study to cosmological distances. First of all, this allows 
 to probe the most powerful objects in the Universe 
(e.g. quasars and AGNs) in which more favorable conditions can be formed for acceleration of protons to
energies $10^{20}$eV. Secondly, because of higher temperature of the 
CMBR at cosmological epochs, i.e.  because of denser and hotter relic photons,
less energetic protons (with energy down to $10^{19}$~eV) can lead to 
effective production of gamma rays. Given that the conditions of acceleration of
protons to $10^{19}$~eV in suspected cosmic accelerators are much relaxed as compared to the $10^{20}$~eV extreme accelerators (see \cite{FAA2002}),
we should expect dramatic increase of such gamma-ray sources. Another factor of
enhancement of number of sources  comes from the increase of the volume of
the explored region to redshifts $z \geq 3$. An obvious caveat in this case
is the decrease of gamma ray flux. However, this factor can be at least
partly compensated by the
huge power of objects in the remote Universe. Moreover, due to dramatic
reduction of mean free paths of UHE protons at cosmological epochs $z \geq 1$, the conversion efficiency of proton energy to gamma rays is
increased almost an order of magnitude, which makes these objects an extremely
effective gamma-ray emitters. Finally,  since the secondary gamma-ray emission
generally follows the direction of parent protons, the beamed cosmic
accelerators like GRBs and blazars seem to be quite  attractive targets for
search of  point-like but steady GeV gamma-ray emission from cosmologically
distant objects.

\appendix  
\section{Angular size of sources at large redshifts and distribution function
in the expanding Universe}\label{app}

If the radiation is generated on scales smaller than cosmological
ones,  the relevant  processes can be considered as they take 
place in the conventional  stationary Euclidean space.  However,
when radiation  propagates over cosmological
distances,  the expansion of space should be taken into account. Based on 
the law of free motion of massless particles in expanding space, one can 
find the distribution function of photons  at  the  observation point. 
Then we need to convert the distribution function  calculated in the 
reference frame with origin at the source to the reference frame of the
observer (\cite{KPA2011}). In the present paper the
free propagation of photons is considered in the flat expanding Universe
with parameters $\Omega_{\Lambda}=0.73$, $\Omega_{m}=0.27$ and $H_{0}=71$
km/s Mpc. 

\begin{figure}[h]
\centering{
\includegraphics[width=0.35\textwidth,angle=-90,clip=]{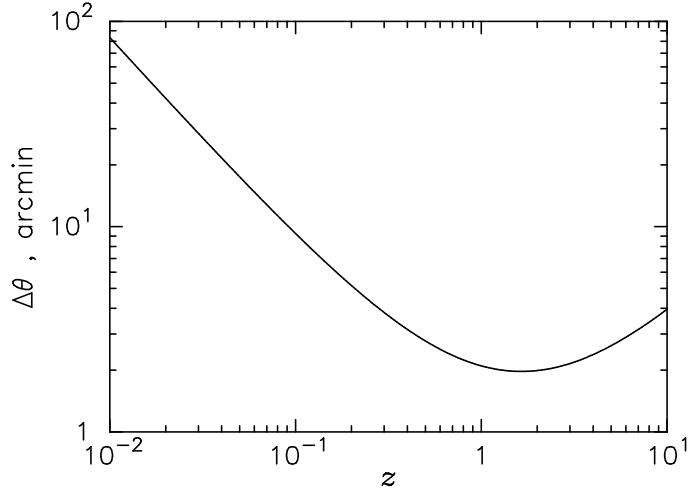}
\caption{\label{fig13} \small Angular size of  the  source with diameter
$D=1$ Mpc located at different redshifts $z$.
 }
}
\end{figure}

We  consider  an  isotropic gamma-ray source of  radius $R_*$ located 
at redshift $z$. Let us assume  that photons   escape this region 
with spherically symmetric distribution $f_{z}(E,\theta)$, where $E$ is the energy 
of photon and $\theta$ is the angle between
photon momentum and radial direction at the escape point.  Finally we assume that 
after the escape gamma rays propagate freely. 
In the case of small angles the distribution function
$f_{0}(z,E,\Omega(\theta))$ of  gamma rays at the observation  point integrated 
over the  solid angle $\Omega$  (with polar angle $\theta$) 
can be expressed in the following form:
\begin{equation}\label{eqa1}
f_{0}(z,E,\Omega(\theta))=2\pi\left(\frac{\Theta_*}{1+z}\right)^2\int\limits_0^{\theta/\Theta_*}f_{z}(E(1+z),x)xdx,
\end{equation}
where      
\begin{equation}\label{eqa2}
\Theta_*=\left(\frac{c}{H_0R_*(1+z)}
\int\limits_0^z\frac{dz'}{\sqrt{\Omega_m(1+z')^3+\Omega_{\Lambda}}}\right)^{-1}.
\end{equation}
The final result (\ref{eqa1}) differs from analogous one corresponding to 
the stationary space by the dependence on the redshift as well as 
by the nonlinear dependence  of angular size on distance. The latter  is 
determined by  the  parameter $\Theta_*$ given by Eq.~(\ref{eqa2}).
The  parameter  $\Delta \theta=2\Theta_*$ has the meaning the angular size 
of the isotropically emitting  source. In the case of anisotropic source, the angular size of radiation cannot be arbitrary large and is limited by $\Delta \theta$.   
In Fig.~\ref{fig13} we show the angular size of the emission region
with diameter $D=2R_*=1$ Mpc. The parameter   $\Delta \theta$
has a minimum $1.95$ arcmin at redshift $z=1.64$ when recession
velocity equals to the speed of light $c$ (\cite{KPA2011}).

\bibliographystyle{aa} 
 
\end{document}